\documentclass[11pt]{article}
\usepackage{moriond,epsfig}

\def\be{\begin{equation}}
\def\ee{\end{equation}}
\def\bea{\begin{eqnarray}}
\def\eea{\end{eqnarray}}

\begin{document}
\vspace*{4cm}
\title{THE CHALLENGE OF DETERMINING SUSY PARAMETERS IN FOCUS-POINT-INSPIRED CASES}

\author{K. ROLBIECKI$^1$\footnote{Speaker, e-mail: Krzysztof.Rolbiecki@fuw.edu.pl},
 K. DESCH$^2$, J. KALINOWSKI$^1$, G. MOORTGAT-PICK$^3$}


\address{$^1$Institute of Theoretical Physics, Warsaw University, PL-00681
Warsaw, Poland\\
$^2$Phys. Institut, Albert-Ludwigs-Universit\"at
Freiburg, D-79104 Freiburg, Germany\\
$^3$TH Division, Physics Department, CERN, CH-1211 Geneva 23,
  Switzerland}

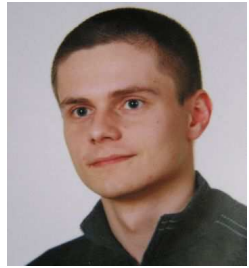
\maketitle\abstracts{We discuss the potential of combined LHC and
ILC experiments for SUSY searches in a difficult region of the
parameter space, in which all sfermion masses are above the TeV scale.
Precision analyses of cross sections of light chargino production
and forward--backward asymmetries of decay leptons and hadrons at the
ILC, together with mass information on ${\tilde{\chi}^0_2}$ and
squarks from the LHC, allow us to fit rather precisely the underlying
fundamental gaugino/higgsino MSSM parameters and to constrain the
masses of the heavy virtual sparticles. For such analyses the
complete spin correlations between the production and decay processes have
to be taken into account. We also took into account expected
experimental uncertainties.}

\section{Introduction}
Supersymmetry (SUSY) is one of the most promising extensions of the
Standard Model (SM) since, among other things, it solves the
hierarchy problem, provides a cold dark matter candidate, and
enables gauge couplings unification. Because of the unknown
mechanism of SUSY breaking, supersymmetric extensions of the
Standard Model contain a large number of new parameters: 105 appear
in the Minimal Supersymmetric Standard Model (MSSM) and have to be
specified. Experiments at future accelerators, the LHC and the ILC,
will have not only to discover SUSY but also to determine precisely
the underlying scenario without theoretical prejudices on the SUSY
breaking mechanism. Particularly challenging are those scenarios
where the scalar SUSY particle sector is heavy, as required e.g.\ in
focus-point scenarios (FP) as well as in split SUSY (sS).

Since it is not easy to determine experimentally cross sections for
production processes, studies have been made to exploit the whole
production-and-decay process. Angular and energy distributions of
the decay products in  production processes with subsequent three-body decays
have been studied for chargino as well as for neutralino processes~\cite{Moortgat-Pick:1998sk}.
Since such observables depend strongly
on the polarization of the decaying particle, the complete spin
correlations between production and decay can have a lot of influence
and must be taken into account. Exploiting such spin effects, it
has been shown~\cite{Moortgat-Pick:1999ck} that, once the
chargino parameters are known, useful indirect bounds for the mass
of the heavy virtual particles could be derived from
forward--backward asymmetries of the final lepton $A_{\rm FB}(\ell)$.

\section{Chosen Scenario: Focus-Point-Inspired Case}

In this section we take a FP-inspired mSUGRA scenario defined at the
GUT scale~\cite{DKMPRS}. However, in order to assess the possibility
of unravelling such a challenging new physics scenario, our analysis
is entirely performed  at the EW scale without any reference to the
underlying SUSY-breaking mechanism. The parameters at the EW scale
are obtained with the help of the code SPheno~\cite{spheno}. The
low-scale gaugino/higgsino/gluino masses, as well as the derived
masses of SUSY particles, are listed in Table~\ref{tab:scenario}. As
can be seen, the chargino/neutralino sector, as well as the gluino,
are rather light, whereas the scalar particles have masses of about
2 TeV (with the only exception of $h$, which is a SM-like light
Higgs boson).

\begin{table}[h]
\begin{center}
\renewcommand{\arraystretch}{1.3}
\begin{tabular}{|c|c|c|c|c|c|c|c|c|c|c|c|}
\hline $M_1$ & $M_2$ & $M_3$ & $\mu$ & $\tan\beta$&
$m_{\tilde{\chi}^{\pm}_1}$ & $m_{\tilde{\chi}^{\pm}_2}$ &
$m_{\tilde{\chi}^{0}_1}$  & $m_{\tilde{\chi}^{0}_2}$ &
$m_{\tilde{\chi}^{0}_3}$ & $m_{\tilde{\chi}^{0}_4}$ &
$m_{\tilde{g}}$
\\ \hline
60 & 121 & 322 & 540 &
 20 & 117 & 552 & 59 & 117 & 545 & 550 & 416 \\ \hline
\hline $m_{h}$ & $m_{H,A}$ & $m_{H^{\pm}}$ & $m_{\tilde{\nu}}$ &
$m_{\tilde{e}_{\rm R}}$ & $m_{\tilde{e}_{\rm L}}$ & $m_{\tilde{\tau}_1}$ &
$m_{\tilde{\tau}_2}$ & $m_{\tilde{q}_{\rm R}}$ & $m_{\tilde{q}_{\rm L}}$ &
$m_{\tilde{t}_1}$ & $m_{\tilde{t}_2}$ \\ \hline 119 & 1934 & 1935 &
1994 & 1996 & 1998 & 1930 & 1963 & 2002 & 2008 & 1093 & 1584 \\
\hline
 \end{tabular}
\caption{Low-scale MSSM parameters and the particle masses in our
scenario in GeV. \label{tab:scenario}}
\end{center}
\end{table}

\subsection{Expectations at the LHC}
As can be seen from Table~\ref{tab:scenario}, all squarks are
kinematically accessible at the LHC. The largest squark production
cross  section is for $\tilde{t}_{1,2}$. However, with stops
decaying mainly to $\tilde{g}t$ [with $BR(\tilde{t}_{1,2}\to
\tilde{g} t)\sim 66\%$], where the background from top production
will be large, the reconstruction of the stops will be very
challenging. The other squarks decay mainly via $\tilde{g} q$, but
since they are very heavy, $m_{\tilde{q}_{\rm L,R}}\sim 2$~TeV,
precise mass reconstruction will be difficult. Nevertheless, the
indication that the scalar quarks are very heavy will be very
important in narrowing experimental uncertainty on the slepton
sector from the ILC measurements.

The gluino production is expected to have very high rates. Therefore
several gluino decay channels can be exploited. The largest
branching ratio for the gluino decay in our scenario is into
neutralinos $BR(\tilde{g}\to \tilde{\chi}^0_2 b \bar{b})\sim 14\%$
with a subsequent leptonic neutralino decay $BR(\tilde{\chi}^0_2\to
\tilde{\chi}^0_1 \ell^+ \ell^-)\sim 6\%$, $\ell=e,\mu$. In this
channel the dilepton edge will be clearly visible, since this process
is practically background-free. The mass difference between the two
light neutralino masses could be measured from the dilepton edge
with an uncertainty of about~\cite{DKMPRS}
\begin{equation}
\delta(m_{\tilde{\chi}^0_2}-m_{\tilde{\chi}^0_1})\sim 0.5~\mathrm{
GeV}. \label{eq-massdiff}
\end{equation}

\subsection{Expectations at the ILC} \label{sec:ilc}
At the ILC with $\sqrt{s}=500$~GeV, only light charginos and
neutralinos are kinematically accessible. However,  in this scenario
the neutralino sector is characterized by very low production cross
sections, below 1~fb, so that it might not be fully exploitable.
Only the chargino pair production process has high rates at the ILC,
see Table~\ref{tab:cross}, and all information obtainable from this
sector has to be used. In the following we study the production
process
\begin{equation}
e^+ e^- \to \tilde{\chi}^+_1 \tilde{\chi}^-_1
\end{equation}
followed by leptonic and hadronic decays of the charginos, for which
the analytical formulae including the complete spin correlations are
given in a compact form~\cite{Moortgat-Pick:1998sk}. The production
process occurs via $\gamma$ and $Z$ exchange in the $s$-channel and
$\tilde{\nu}_e$ exchange in the $t$-channel, and the decay processes
get contributions from $W^{\pm}$ and $\tilde{\nu}_\ell$,
$\tilde{\ell}_{\rm L}$ (leptonic decays) or $\tilde{q}_{d_{\rm L}}$,
$\tilde{q}_{u_{\rm L}}$ exchange (hadronic decays). The light
chargino has a leptonic branching ratio of about
$BR(\tilde{\chi}^{-}_1\to \tilde{\chi}^0_1 \ell^-
\bar{\nu}_\ell)\sim 11\%$ for each family  and a hadronic branching
ratio of about $BR(\tilde{\chi}^-_1\to \tilde{\chi}^0_1 q_d
\bar{q_u})\sim 33\%$.

In our analysis we use cross sections multiplied by the branching
ratios of semileptonic chargino decays:
$\sigma(e^+e^-\to\tilde{\chi}^+_1\tilde{\chi}^-_1)\times BR$, with
$BR=2\times BR(\tilde{\chi}^+_1 \to \tilde{\chi}^0_1 \bar{q}_d
q_u)\times BR(\tilde{\chi}^-_1 \to \tilde{\chi}^0_1 \ell^-
\bar{\nu}_\ell)+ [BR(\tilde{\chi}^-_1 \to \tilde{\chi}^0_1 \ell^-
\bar{\nu}_\ell)]^2\sim 0.34$, $\ell=e,\mu $, $q_u=u,c$, $q_d=d,s$.

\begin{table}
\begin{center}
\renewcommand{\arraystretch}{1.3}
\begin{tabular}{|l c||c||c|} \hline $\sqrt{s}$/GeV &
$(P_{e^-},P_{e^+})$ &
$\sigma(\tilde{\chi}^+_1\tilde{\chi}^-_1)\times BR\times
\varepsilon$/fb
& $A_{FB}(\ell^-)$/\% \\
\hline\hline 350 & $(-90\%,+60\%)$ & 1062.5$\pm$4.0
& {4.42$\pm$0.29} \\
\hline & $(+90\%,-60\%)$ & 14.6$\pm 0.7$ & 4.6$\pm 2.5$\\
\hline\hline 500 & $(-90\%,+60\%)$ & 521.6$\pm
2.3$ & {4.62$\pm$0.41}\\
\hline & $(+90\%,-60\%)$ & 6.9$\pm 0.4$ & 4.9$\pm 3.6$\\
\hline
\end{tabular}
\end{center}
\caption{Cross sections for the process $e^+e^-\to
\tilde{\chi}^+_1\tilde{\chi}^-_1$ and forward--backward asymmetries
for this process followed by $\tilde{\chi}^-_1 \to \tilde{\chi}^0_1
\ell^- \bar{\nu}_{\ell}$, with $\ell=e,\mu$, for different beam
polarization $P_{e^-}$, $P_{e^+}$ configurations at the center of
mass energies $\sqrt{s}=350$~GeV and $500$~GeV at the ILC. Errors
include $1\sigma$ statistical uncertainty assuming luminosity ${\cal
L}=200$~fb$^{-1}$ for each polarization configuration, efficiency
$\varepsilon=50\%$ and the beam polarization uncertainty of~0.5\%.
$BR\simeq 0.34$, see Section \ref{sec:ilc}. \label{tab:cross}}
\end{table}

From the ILC scan at the threshold~\cite{stirling}, because of the steep
$s$-wave excitation curve in $\tilde{\chi}^+_1\tilde{\chi}^-_1$
production, the determination of the light chargino mass will be
possible with an accuracy of about~\cite{TDR}
\begin{equation}
m_{\tilde{\chi}^{\pm}_1}=117.1 \pm 0.1 ~\mathrm{ GeV}.
\label{eq_massthres}
\end{equation}
The mass of the lightest neutralino $m_{\tilde{\chi}^0_1}$ can be
derived either from the energy distribution of the lepton $\ell^-$
or, in  hadronic decays, from the invariant mass distribution of the
two jets from $\tilde{\chi}^\pm_1$ decays. We therefore
assume~\cite{DKMPRS} that
\begin{equation}
m_{\tilde{\chi}^{0}_1}=59.2 \pm 0.2 ~\mathrm{ GeV}.
\label{eq_masslsp}
\end{equation}
Together with the information from the LHC, Eq.~(\ref{eq-massdiff}),
a mass uncertainty for the second lightest neutralino of about
\begin{equation}
m_{\tilde{\chi}^{0}_2}=117.1 \pm 0.5 ~\mathrm{ GeV}
\label{eq_masschi02}
\end{equation}
can be assumed.

\section{Determination of Parameters}
Following the method proposed in~\cite{Moortgat-Pick:1999ck} we
include in the fit the forward--backward asymmetries of the final
leptons. As explained before, this observable is very sensitive to
the mass of the exchanged scalar particles, even for rather heavy
masses. Since in the decay process also the left selectron exchange
contributes, the $SU(2)$ relation between the left selectron and
sneutrino masses: $m^2_{\tilde{e}_{\rm L}}= m_{\tilde{\nu}_e}^2 + m_Z^2
\cos(2 \beta) (-1+\sin^2\theta_W)$ has been assumed. In principle
this assumption could also be relaxed~\cite{DKMPRS}.

Applying the 5-parameter $\chi^2$ fit procedure with the leptonic
forward--backward asymmetries included leads to~\cite{DKMPRS}:
\begin{eqnarray}
&&M_1=60.00\pm 0.35~\mathrm{ GeV},\quad M_2=121.0\pm 1.1~\mathrm{
GeV},\quad 500\leq\mu \leq 610~\mathrm{ GeV},\nonumber \\ &&
m_{\tilde{\nu}_e}=1995\pm 100~\mathrm{ GeV},\quad 14\leq \tan\beta
\leq 31.
\end{eqnarray}
Including forward--backward asymmetries in the multiparameter fit
provides strong constraints for the mass of the heavy virtual
particle, $m_{\tilde{\nu}_e}$, and decreases its error by a factor
of about $2$ with respect to the fit without FB
asymmetry~\cite{DKMPRS}. The constraints for the gaugino mass
parameters $M_1$ and $M_2$ are improved by a factor of about $5$,
thanks to the constraint on the value of $\tan\beta$. It is clear
that in order to improve considerably the bounds for the parameters
$\mu$ and $\tan\beta$, the measurement of the heavy higgsino-like
chargino and/or neutralino masses will be necessary at the second
phase of the ILC with $\sqrt{s}\sim 1000$ GeV.

\section{Conclusions}
We have demonstrated a method for constraining heavy virtual
particles and for determining the SUSY parameters in focus-point-inspired
scenarios. These appear very challenging since
only  little experimental information on the SUSY
sector is accessible at both the LHC and the ILC at its first energy
stage of $\sqrt{s}=500$ GeV. However, we show that a careful
exploitation of the data leads to significant constraints on the unknown
parameters. The most powerful tool in this kind of analysis turns
out to be the forward--backward asymmetry. A proper treatment of
spin correlations between the production and the decay is indispensable in
that context. This asymmetry is strongly dependent on the mass of
the exchanged heavy particle. We want to stress the important role
of the LHC/ILC interplay since none of these colliders alone can
provide us with the data needed to perform the determination of the SUSY parameters
in focus-like scenarios.

\section*{Acknowledgements}
The author would like to thank the organizers for financial support
and the opportunity to give this talk. JK and KR have been supported
by the Polish Ministry of Science and Higher Education Grant No 1
P03B 108 30.

\end{document}